\documentclass[11pt]{article}
\usepackage[numbers,sort&compress]{natbib}
\usepackage{amsmath}
\usepackage{amssymb}
\newcommand{\B}{{\cal B}}

\newcommand{\C}{{\cal C}}
\newcommand{\F}{{\cal F}}

\newtheorem{theorem}{Theorem}[section]

\newtheorem{lemma}[theorem]{Lemma}

\newtheorem{example}[theorem]{Example}

\def\whitebox{{\hbox{\hskip 1pt
 \vrule height 6pt depth 1.5pt
 \lower 1.5pt\vbox to 7.5pt{\hrule width
    3.2pt\vfill\hrule width 3.2pt}%
 \vrule height 6pt depth 1.5pt
 \hskip 1pt } }}
\def\qed{\ifhmode\allowbreak\else\nobreak\fi\hfill\quad\nobreak
     \whitebox\medbreak}

\newcommand{\ignore}[1]{}
\begin{document}

\baselineskip 16pt
\title{Constructions and bounds for separating hash families}
\author{\small  X. Niu and H. Cao \thanks{Research supported by the National Natural
Science Foundation of China under Grant No. 11571179,
and the Priority Academic Program Development of Jiangsu Higher
Education Institution.
} \\
\small Institute of Mathematics, \\ \small  Nanjing Normal
University, Nanjing 210023, China}

\date{}
\maketitle

\begin{abstract}
In this paper, we present a new construction for strong separating hash families by using hypergraphs and obtain some optimal separating hash families.
 We  also improve some previously known
bounds of separating hash families.

\bigskip

\noindent {\textbf{Key words:}} Hash family, separating hash family, strong separating hash family, hypergraph

\bigskip

\end{abstract}

\section{Introduction}

Let $X$ and $Y$ be two finite sets of sizes $n$ and $m$ respectively. An $(N;n,m)$-{\it hash family} $\F$ is a family
of functions from $X$ to $Y$ with $|\F| = N$.
 For all pairwise disjoint subsets $ C_1,C_2,\ldots, C_t \subseteq X $,
if there exists some $f$ such that $ f(C_i)\cap f(C_j)= \emptyset $ for all $1\le i < j\le t$.
Then $C_1,C_2,\ldots,C_t$ are {\it separable} in $\F$, and the function $f$ is said to
{\it separate} the sets $C_1,C_2,\ldots,C_t$.

 Given positive
integers $w_1,w_2,\ldots, w_t$, we say  $\F$ is a $
\{w_1,w_2,\ldots, w_t \}$-{\it separating hash family}, denoted by
SHF$(N;n,m,\{w_1,w_2,\ldots, w_t \})$, if for all pairwise
disjoint subsets $ C_1,C_2,\ldots, C_t \subseteq X $ with $|C_i| =
w_i$ for $i=1,2,\ldots,t$, there exists some $f \in \F$ which separates $ C_1,C_2,\ldots, C_t $.
The parameter multiset
$\{w_1,w_2,\ldots, w_t \}$ is called the {\it type} of  $\F$.
For the sake of brevity, we use SHF to denote separating hash family,
and we also use $\{w_1^{q_1},w_{2}^{q_{2}},\ldots,w_t^{q_t}\}$  to denote the
multiset in which there are exactly $q_i$ copies of $w_i$ and
$w_i<w_j$ for $1\le i<j\le t$. Further, $w^1$ will be written as
$w$. An SHF$(N;n,m,\{1^q, w\})$ with $w\ge 2$ is also called a {\it strong separating hash family}.

Separating hash families were first introduced by Stinson, Trung
and Wei \cite{STW}. It can be used to construct frameproof codes,
secure frameproof codes and parent-identifying codes, see
\cite{BS,SS,SSW,STW}. Most results of the known
papers on separating hash families are focused on their bounds and
constructions, see \cite{AS,AMSW,BT0,BT,BSR,BESZ,RFH,GG,GSW,GSW,GST,GS,SGWGM,SWC,WC}.

Given an SHF$(N;n,m,\{w_1,w_2,\ldots, w_t \})$, we
 construct an $N\times n$ matrix $A=(a_{i,j})$ having entries on
a set of $m$ elements such that $a_{i,j}=f_i(x_j)$ where
 $f_1,f_2,\ldots,f_N$ are some fixed ordering of the functions in $\F$ and $x_1,\ldots,x_n$ are elements of $X$.
 This matrix is called the {\it representation matrix} of the SHF.
For all disjoint sets of columns $ C_1,C_2,\ldots, C_t $ of $A$ with $ |C_i|= w_i$, $1\le i\le t$,
there exists at least one row $r$ of $A$ such that
$\{a_{r,x}: x\in C_i\}\cap \{a_{r,y}: y\in C_j\}= \emptyset $ holds for all
$1\le i<j\le t$. We say the row $r$ separates the sets $ C_1,C_2,\ldots, C_t $.
An  SHF$(N;n,m,\{w_1,w_2,\cdots,w_t\})$ is  called {\it optimal}  if $n$ is maximum
for given $N, m,w_1,w_2,\cdots,w_t$ or if $N$ is minimum
for given $n, m,w_1,w_2,\cdots,w_t$.

In the literature optimal results for separating hash families are quite rare.
In this paper, we  present a new construction for strong separating hash families by using hypergraphs
and obtain some optimal separating hash families.
It is easy to see that a separating hash family with type $\{w_1,w_2,\ldots,w_t\}$ is
 a  separating hash family with type $\{w_1+w_2, w_3,\ldots,w_t\}$ and also
 a  separating hash family with type $\{w, w_2, \ldots,w_t\}$ with $w\le w_1$.
So separating hash families with type $\{w_1,w_2, \ldots,w_t\}$ can yield separating hash families with type $\{w_1',w_2'\}$.
Thus it is valuable to study the bound of separating hash families with type $\{w_1,w_2\}$.

This paper is organized as follows. In the next section,
we present a new construction for strong separating hash families by using hypergraphs
and  present some optimal strong separating hash families.
In Section 3 we construct an optimal SHF$(4;10,4,\{2,2\})$ and use it to improve the  known
bound for an SHF$(2w;n,m,\{w,w\})$.
In the last section, we  improve the  known
bound for an SHF$(w_1+w_2;n,m,\{w_1,w_2\})$ and use it to improve the  known
bound for an SHF$(\sum_{i=1}^{t}w_i;n,m,\{w_1,w_2,\ldots,w_t\})$.

\section{A new construction for strong separating hash family }

In this section, we will give a construction for strong separating hash families by using hypergraphs,
and we also give some tight results for strong separating hash families.
A {\it hypergraph} is a pair $H = (V,E)$, where $V$ is a finite set whose elements are called
vertices and $E$ is a family of subsets of $V$, called edges. It is $k$-uniform if each of
its edges contains precisely $k$ vertices.

\begin{theorem}\label{cons}
Let $V= \{ x_1, x_2, \ldots, x_n  \}$ and $E=\{ B_1, B_2,\ldots, B_N \}$, where
 $B_{i}$ is an $m$-subset of $V$, $1\le i\le N$.
  If $G=(V,E)$ is an $m$-uniform hypergraph
with the property that any $l$ vertices are contained in exactly one edge,
then there exists an {\rm SHF}$(N;n,m+1,\{1^{w_1},w_2\})$ for all positive integers $w_1$ and $w_2$ satisfying $w_1\le l$ and $w_1+w_2\le n$.
\end{theorem}

\noindent {\it Proof:} Let $B_{i}=\{y_{i,1},y_{i,2},\ldots,y_{i,m}\}, 1\le i\le N$. Define an $N \times n$  matrix $A= (a_{i,j})$ by
$$a_{i,j}= \begin{cases}
 b, & \text{ if }  ~x_j=y_{i,b}\\
 0, & \text{ otherwise.}
 \end{cases}$$
Now we  prove  $A$ is a  representation  matrix of an
SHF$(N;n,m+1,\{1^{w_1},w_2\})$.
Let $\C=\{c_1,c_2,\dots,c_n\}$ denote the column set of  $A$.
Let $ C_1, C_{2}, \dots, C_{w_1},C_{w_1+1}$ be pairwise disjoint subsets of $\C$ such that
$C_{i}=\{c_{s_i}\}$ for $i=1,2, \ldots, w_1$ and $\mid C_{w_1+1}\mid =w_2\le n-w_1$.
Then $\mid C_1\cup C_{2}\cup  \dots \cup C_{w_1}\mid=w_1\le l$, $C_{w_1+1}\subset\C\backslash \{c_{s_1},c_{s_2},\dots, c_{s_{w_1}}\}$.

Since $G=(V,E)$ is an $m$-uniform hypergraph with the property
that any $l$ vertices are contained in exactly one edge,
we can find an edge $B_t\in E$ such that
$\{x_{s_1}, x_{s_2}, \dots, x_{s_{w_1}}\}\subset B_{t}$. Let $B_{t}=\{x_{s_1}, x_{s_2}, \dots, x_{s_{w_1}},x_{s_{w_1+1}},\dots,x_{s_{m}}\}$.
Then $\{a_{t,s_1},a_{t,s_2},\dots,a_{t,s_{m}}\}=\{1,2,\dots,m\}$ and $a_{t,j}=0$ for any $j\in \{1,2,\dots,n\}\backslash \{s_1,s_2,\dots,s_{m}\}$.
Thus, $\{a_{t,k}: c_k\in C_i\}\cap \{a_{t,k}: c_k\in C_j\}= \emptyset $ holds for all
$1\le i<j\le w_1+1$.
Therefore,  the $t$-th row of $A$ can separate  $ C_1,C_2,\ldots, C_{w_1+1} $.\qed

\begin{example} Let  $V= \mathbb{Z}_{7}$ and $E=\{\{i, i+1, i+3\}: i\in \mathbb{Z}_{7}  \}$. We can obtain the following representation  matrix of an {\rm SHF}$(7;7,4,\{1^{2},5\})$ by Theorem~\ref{cons}.

\begin{center}
\begin{tabular}{|c|c|c|c|c|c|c|}
  \hline
  $1$     & $2$      & $0$   & $3$       & $0$       & $0$   & $0$        \\
  \hline
  $0$     & $1$     & $2$      & $0$   & $3$       & $0$       & $0$      \\
  \hline
  $0$     & $0$     & $1$     & $2$      & $0$   & $3$       & $0$           \\
  \hline
  $0$     & $0$      & $0$  & $1$     & $2$      & $0$   & $3$         \\
  \hline
   $3$    & $0$     & $0$      & $0$  & $1$     & $2$      & $0$           \\
  \hline
   $0$    & $3$    & $0$     & $0$      & $0$  & $1$     & $2$                \\
  \hline
   $2$    &$0$    & $3$    & $0$     & $0$      & $0$  & $1$                   \\
  \hline
\end{tabular}
\end{center}
\end{example}

For our results, we need the following conclusion on  $m$-uniform hypergraphs.

\begin{theorem}{\rm (\cite{HH1960,HH1975,HDR1965})}\label{sshf-c}
 1.  For any $3\le m \le 5$,
there exists an  $m$-uniform hypergraph $G=(V,E)$ with the property that
 any two vertices are contained in exactly one edge, where $|V|=n$ satisfying $n\equiv 1, m \pmod {m^2-m}$, and $|E|=\frac{n(n-1)}{m(m-1)}$.

\noindent 2. For any $n\equiv 2, 4 \pmod {6}$,
there exists a  $4$-uniform hypergraph with $n$ vertices and $\frac{n(n-1)(n-2)}{24}$ edges such that
 any three  vertices are contained in exactly one edge.

\noindent 3. For any prime power $m$ and integer $l\ge2$,
there exists an  $(m+1)$-uniform hypergraph $G=(V,E)$ with the property that
 any three vertices are contained in exactly one edge, where $|V|=n$ satisfying $n=m^l+1$, and $|E|=\frac{\binom{n}{3}}{\binom{m+1}{3}}$.
 \end{theorem}

 By Theorems~\ref{cons} and \ref{sshf-c} we have the following theorem.

\begin{theorem}\label{sshf-2}
 1.  Let $3\le m \le 5$, $n\equiv 1, m \pmod {m^2-m}$, $n> m$, and $N=\frac{n(n-1)}{m(m-1)}$.
Then there is an {\rm SHF}$(N;n,m+1,\{1^{2},n-2\})$.

\noindent 2. Let $n\equiv 2, 4 \pmod {6}$, $n\ge 8$ and $N= \frac{n(n-1)(n-2)}{24}$. Then there is an {\rm SHF}$(N;n,5,\{1^{3},n-3\})$.

\noindent 3.  Let $n=m^l+1$, where $m$ is a prime power and $l\ge 2$. Let $N=\frac{\binom{n}{3}}{\binom{m+1}{3}}$.
Then there is an {\rm SHF}$(N;n,m+2,\{1^{3},n-3\})$.

\end{theorem}

Now we have obtained some  new  strong separating hash families from  Theorem \ref{sshf-2}.
We continue to show that these results from Theorem~\ref{sshf-2} are all tight by discussing the lower bound of $N$ for an SHF$(N;w_1+w_2,m,\{1^{w_1},w_2\})$.
When $w_1+w_2\le m$, it is easy to see that $N\ge 1$. So this case is trivial and we only need to deal with the case  $w_1+w_2>m$.

\begin{theorem}\label{sshf-1}
If there is an {\rm SHF}$(N;w_1+w_2,m,\{1^{w_1},w_2\})$ with $w_1+w_2>m$,
then $N\ge \frac{\binom{w_1+w_2}{w_1}}{\binom{m-1}{w_1}}$.
\end{theorem}

\noindent {\it Proof:}  Let $A=(a_{i,j})$ be a representation matrix of an SHF$(N;w_1+w_2,m,\{1^{w_1},w_2\})$
with the column set $\C=\{c_1,c_2,\dots,c_n\}$.
Then the total number of the pairwise disjoint subsets of $\C$ which need to be separated is $\binom{w_1+w_2}{w_1}$.
On the other hand, suppose $C_1, C_{2}, \ldots, C_{w_1+1}$ are pairwise disjoint subsets of $\C$ such that
$\mid C_{w_1+1}\mid =w_2$ and $\mid C_{j}\mid =1$ for
$1\le j\le w_1$. Let $C_{j}=\{c_{i_j}\}$ for $1\le j\le w_1$.
  If these $w_1+1$ subsets can be separated by the $i$th row,  then the element $a_{i,j}$ appears exactly once in the $i$th row for any $j\in\{i_1,i_2,\dots,i_{w_1}\}$.
Since $w_1+w_2>m$, there are at most $m-1$ elements occurring exactly once in each row.
So the maximum number of the pairwise disjoint subsets which can be separated by each row is $\binom{m-1}{w_1}$.
Thus,  $N\ge \frac{\binom{w_1+w_2}{w_1}}{\binom{m-1}{w_1}}$.\qed

By Theorem~\ref{sshf-1}, it is easy to check that these  strong separating hash families constructed in  Theorem~\ref{sshf-2} are all optimal.
For more results on $m$-uniform hypergraphs with the property that
 any $t$ vertices are contained in exactly one edge, see \cite{CC}(pages 72-73, 82-84, 661).
For example,  there are a $5$-uniform hypergraph with 11 vertices and 66 edges  such that any four vertices are contained in exactly one edge,
and  a $6$-uniform hypergraph with 12 vertices and 132 edges  such that any five vertices are contained in exactly one edge, see \cite{CC}(page 661).
 By Theorems  \ref{cons} and \ref{sshf-1}, we obtain an optimal SHF$(66;11,5,\{1^{4},7\})$ and an optimal SHF$(132;12,6,\{1^5,7\})$.

\noindent \textbf{Remark 1 :} For the bounds of strong separating hash families, Sarkar and Stinson \cite{SS} proved that there exists an infinite class of SHF$(N;n,m,\{1^{w_1}, w_2\})$ for which  $N$ is $O((w_1(w_1+w_2))^{log^*n}log~ n)$.
Liu and Shen \cite{LS} gave an infinite constructions of the SHF$(N;n,m,\{1^{w_1}, w_2\})$ for which  $N$ is $O(log~ n)$.
Guo and Stinson \cite{GS} proved  $ N \ge \binom{n}{m-1}$ when $w_1\ge m-1$
 and $w_1+w_2\le n\le 2(w_1+w_2)-m $. Now we compare our conclusion with the bound in \cite{GS}.
 By the definition of an SHF, it is obvious that $m-1\ge w_1$. So Guo and Stinson's bound in \cite{GS} can be restated as $ N \ge \binom{n}{w_1}$
when $m-1=w_1$ and $w_1+w_2\le n\le w_1+2w_2-1$. When $n=w_1 +w_2$ and $m-1=w_1$, it is easy to see that we have the same conclusion.
But for the case $n=w_1 +w_2$ and $m-1>w_1$, we have a new tight bound.

Actually, the hypergraphs used in Theorem~\ref{cons} need not to be $k$-uniform, and any $l$ vertices need not to be contained in exactly one edge.
We can generalize this construction to a hypergraph with different edge sizes.
The proof of the following theorem is similar to that of Theorem~\ref{cons}. So we just present the theorem without proof.

\begin{theorem}\label{cons-1}
Suppose there is a hypergraph with $n$ vertices and $N$ edges such
that the maximum edge size is $m$ and any $l$ vertices are contained in at least one edge,
then there exists an {\rm SHF}$(N;n,m+1,\{1^{w_1},w_2\})$ for all positive integers $w_1$ and $w_2$ satisfying $w_1\le l$ and $w_1+w_2\le n$.
\end{theorem}

Below are some results of  $k$-uniform hypergraphs such that any $l$ vertices are contained in at least one edge.

\begin{theorem}{\rm (\cite{CC,CY})}\label{sshf-c3}
 1.There exists an  $3$-uniform hypergraph $G=(V,E)$ with the property that
 any two vertices are contained in at least one edge, where $|V|=n\ge 5$  and $|E|=\lceil\frac{n}{3}\lceil\frac{n-1}{2}\rceil\rceil$.

\noindent 2.There exists an  $4$-uniform hypergraph $G=(V,E)$ with the property that
 any two vertices are contained in at least one edge, where $|V|=n\ge 6$  and
 $|E|=\lceil\frac{n}{4}\lceil\frac{n-1}{3}\rceil\rceil + c$, where $c=1$ if $n=7,9,10$; $c=2$ if $n=19$; and $c=0$ for all other values of $n$.

\noindent 3. There exists an  $4$-uniform hypergraph $G=(V,E)$ with the property that
 any three vertices are contained in at least one edge, where $|V|=n$ and $|E|=\lceil\frac{n}{4}\lceil\frac{n-1}{3}\lceil\frac{n-2}{2}\rceil\rceil\rceil$ satisfying $n\neq 12t+7$ for any
 $t\in([0, 12]\backslash \{6\})\cup \{ 16, 21, 23, 25, 29\}$.
\end{theorem}

For more results on $m$-uniform hypergraphs with the property that
 any $t$ vertices are contained in at least one edge, see \cite{CC}(pages 366-372).
By Theorems~\ref{cons-1} and  \ref{sshf-c3} we have the following theorem.

\begin{theorem}\label{sshf-3}
1. For $n\ge 5$, there is an {\rm SHF}$(N;n,4,\{1^{2},n-2\})$ with $N = \lceil\frac{n}{3}\lceil\frac{n-1}{2}\rceil\rceil$.

\noindent 2. For $n=7,9,10$, there is an {\rm SHF}$(N;n,5,\{1^{2},n-2\})$ with $ N=\lceil\frac{n}{4}\lceil\frac{n-1}{3}\rceil\rceil+1$.

\noindent 3.  There is an {\rm SHF}$(31;19,5,\{1^{2},17\})$.

\noindent 4.  For $n \ge 6$ and $n\neq 7,9,10,19$, there is an {\rm SHF}$(N;n,5,\{1^{2},n-2\})$ with $ N=\lceil\frac{n}{4}\lceil\frac{n-1}{3}\rceil\rceil$.

\noindent 5. For any $n\neq 12t+7$, $t\in([0, 12]\backslash \{6\})\cup \{ 16, 21, 23, 25, 29\}$,
there is an {\rm SHF}$(N;n,5,\{1^{3},n-3\})$ with $N = \lceil\frac{n}{4}\lceil\frac{n-1}{3}\lceil\frac{n-2}{2}\rceil\rceil\rceil$.

\end{theorem}

By Theorem \ref{cons-1}, we can obtain more results on strong separating hash families by using the known results on hypergraphs.
For example, there exists a hypergraph with 10 vertices and 12 edges with size 3 or 4 such that any two vertices are contained in exactly one edge, see \cite{CC}(page 231).
 Thus, we can construct an SHF$(12;10,5,\{1^{2},8\})$.
 There  exists a hypergraph with 16 vertices and 68 edges with size 4 or 5 such that any three vertices are contained in exactly one edge, see \cite{CC}(page 662).
 Thus, we can construct an SHF$(68;16,6,\{1^{3},13\})$.
There exists a hypergraph with 17 vertices and 252 edges with size 5 or 6 such that any four vertices are contained in exactly one edge, see \cite{CC}(page 663).
 Thus, we can construct an SHF$(252;17,7,\{1^{4},13\})$.
There also exists a hypergraph with 16 vertices and 478 edges with size 6 or 8 such that any five vertices are contained in exactly one edge, see \cite{CC}(page 661).
 Thus, we can construct an SHF$(478;16,9,\{1^{5},11\})$.
For more results of the known papers on hypergraphs, see \cite{CC}.

 For given parameters $n,m,w_1,w_2$ of an SHF$(N;n,m+1,\{1^{w_1},w_2\})$, we can also obtain the following  bounds on $N$. Let $M(n, k, l)$ denote the minimum possible number of edges of a $k$-uniform hypergraph with the property
that any $l$ vertices are contained in at least one edge.

\begin{theorem}{\rm (\cite{VR})}\label{sshf-c2}
 $\frac{\binom{n}{l}}{\binom{k}{l}}\le  M(n, k, l)\le (1+o(1))\frac{\binom{n}{l}}{\binom{k}{l}}$,
 where the $o(1)$ term tends to zero as $n$ tends to infinity.
\end{theorem}

According to  Theorems  \ref{cons-1} and \ref{sshf-c2}, we have the following  theorem.

\begin{theorem}\label{sshf-4}
If there exists an SHF$(N;n,m+1,\{1^{w_1},w_2\})$ with $w_1\le m$, then $N\le (1+o(1))\frac{\binom{n}{w_1}}{\binom{m}{w_1}}$.
Furthermore, if $n= w_1 +w_2$, we have
$\frac{\binom{n}{w_1}}{\binom{m}{w_1}} \le N\le (1+o(1))\frac{\binom{n}{w_1}}{\binom{m}{w_1}}$.
\end{theorem}

\section{An improved bound for SHF$(2w;n,m,\{w,w\})$}

In this section, we shall give a new bound for an
SHF$(2w;n,m,\{w,w\})$ with $w\ge 2$.
This bound is useful in the next section.
We start with some definitions and notations.

Let $A$ and $B$ be two matrices.  If $B$ can be
obtained from $A$ by permuting the rows and/or columns
and/or elements, then we say that $A$ is {\it isomorphic} to $B$.

 \begin{lemma}\label{r-c}
 If $A$ is a representation  matrix of an
{\rm SHF}$(N;n,m,\{w_1,w_2,\ldots,w_t \})$, and if $B$
is isomorphic to $A$, then $B$ is also a representation
 matrix of an {\rm SHF}$(N;n,m,\{w_1,w_2,\ldots,w_t\})$.
 \end{lemma}

In order to find good upper bounds for SHFs,
we often use one of the general methods to show that a particular choice of $n$ implies that the representation
 matrix $A$ always contains a submatrix which is impossible in an SHF with given parameters.
 Such a submatrix is referred to a {\it forbidden configuration}.

\begin{lemma}
\label{1.9}  If $A$ is a representation matrix of an
{\rm SHF}$(4;n,m,\{2,2\})$, then  any
submatrix of $A$ can't be isomorphic to the following forbidden configurations $F_1$, $F_2$ and $F_3$.

\begin{center}
\begin{tabular}{|c|c|c|c|}
\multicolumn{4}{c}{$F_1$}\\
  \hline
  $a$  & $a$ & $\ast$   & $\ast$ \\
 \hline
  $b$  & $b$ & $\ast$   & $\ast$ \\
 \hline
   $\ast$ & $\ast$ & $c$ & $c$ \\
 \hline
   $\ast$ & $\ast$ & $d$ & $d$ \\
 \hline
\end{tabular}
\hspace{1cm} \begin{tabular}{|c|c|c|c|}
\multicolumn{4}{c}{$F_2$}\\
  \hline
  $a$  & $a$ & $\ast$    & $\ast$ \\
 \hline
  $b$  & $b$ & $\ast$    & $\ast$ \\
 \hline
   $x$ & $\ast$ & $x$  & $\ast$ \\
 \hline
   $\ast$ & $v$ & $\ast$    & $v$ \\
 \hline
\end{tabular}
\hspace{1cm}
\begin{tabular}{|c|c|c|c|}
\multicolumn{4}{c}{$F_3$}\\
  \hline
  $a$   & $a$     & $\ast$ & $\ast$ \\
 \hline
 $\ast$ & $b$     & $b$    & $\ast$ \\
 \hline
 $x$    & $\ast$  & $\ast$    & $x$    \\
 \hline
 $\ast$    & $\ast$  & $v$    & $v$    \\
 \hline
\end{tabular}
\end{center}
\end{lemma}

\noindent {\it Proof:} It is easy to check that in  $F_1$ or $F_3$ the
column sets $C_1 = \{1,3\}$ and $C_2 =\{2,4\}$ are not
separable, and  in  $F_2$ the column sets $C_1 = \{1,4\}$ and
$C_2 =\{2,3\}$ are not separable. \qed

Suppose $A=(a_{i,j})$ is an $N\times n$ representation matrix of an
SHF on $m$ elements in $Y$. We need the following notations for the following lemmas. Let
$$\lambda_x^i=|\{j: a_{i,j}=x, 1\le j\le n\}|, \ 1\le i\le N,
\ x\in Y,$$ $$\lambda_{max}=max\{\lambda_{x}^i:1\le i\le N,
x\in Y\},\ \mbox{and}$$ $$d_{i,j}(x,y)=|\{k :a_{i,k}=x,a_{j,k}=y, 1\le
k\le n\}|,\ 1\le i<j\le N.$$
Suppose $B$ is a submatrix of $A$, we shall use $A-B$ to denote the matrix obtained by removing all these entries in $B$ from $A$.

\begin{lemma}
\label{aabb} Suppose $A$ is a representation matrix of an
{\rm SHF}$(4;n,m,\{2,2\})$ with $m \geq3 $. If there is a pair of
elements $x$ and $y$ such that $d_{i,j}(x,y)\ge 2$, then $ n\leq
(m-1)^2 +1$.
\end{lemma}

\noindent {\it Proof:} Without loss of generality, we may assume
$d_{1,2}(a,b)\ge 2$. Then the matrix $B$ in Table 1 is a submatrix of $A$.

\begin{center}
{\bf Table 1: Three submatrices of $A$}

$B=$ \begin{tabular}{|c|c|}
  \hline
  $a$   & $a$  \\
 \hline
  $b$   & $b$ \\
 \hline
   $x$  & $y$ \\
 \hline
   $u$  & $v$  \\
 \hline
\end{tabular}
\hspace{1cm}
$E=$ \begin{tabular}{|c|c|c|}
\hline
  $a$ & $a$ & $\ast$  \\
 \hline

  $b$ & $b$ & $\ast$  \\
 \hline

   $x$ & $y$ & $y$  \\
 \hline

   $u$ & $v$ & $u$ \\

 \hline

\end{tabular}
\hspace{1cm}
$M =$ \begin{tabular}{|c|c|c|c|}

  \hline
  $a$ & $a$ & $\ast$ & $\ast$ \\
 \hline

  $b$ & $b$ & $\ast$ & $\ast$ \\
 \hline

   $x$ & $y$ & $y$ & $x$  \\
 \hline

   $u$ & $v$ & $u$ & $v$\\

 \hline

\end{tabular}
\end{center}

 By Lemma~\ref{r-c} we only need to consider the following 5 cases.

1. $\lambda_x^3=\lambda_y^3=1$. Then $A-B$ is a representation matrix of an
SHF$(4;n-2,m,\{2,2\})$, and there are at most $m-2$ distinct
elements  in row three. By the pigeonhole principle, there is an
element $t$ in row four such that $\lambda_t^4\ge \lceil \frac{n-2}{m}\rceil$.
If $\lambda_t^4>m-2$, then we have an element $g$ in  row three such that $ d_{3,4}(g,t)\ge 2$ by the pigeonhole principle.
So we obtain a submatrix of $A$ which is isomorphic to the forbidden configuration $F_1$.
Thus, $\lambda_t^4\le m-2$.
Then  we have $\lceil \frac{n-2}{m}\rceil \leq m-2$.
So, $n\leq m^2 -2m +2$.

2. $\lambda_x^3=1$, $\lambda_y^3>1$, $\lambda_u^4=1$,
$\lambda_v^4>1$. Let $D$ be the $4\times (n-1)$ matrix obtained
from $A$ by removing column one. Then $D$ is a representation
matrix of an SHF$(4;n-1,m,\{2,2\})$, and there are at most
 $m-1$ distinct elements  in the last two rows. By the pigeon
hole principle, there is an element $t$ such that $\lambda_t^4\ge
\lceil \frac{n-1}{m-1}\rceil$. Then we have $\lceil
\frac{n-1}{m-1}\rceil \leq m-1$ and $n\leq m^2 -2m +2$.

3. $\lambda_x^3=1$, $\lambda_y^3>1$, $\lambda_u^4>1$,
$\lambda_v^4=1$.  By Lemma~\ref{1.9}( forbidden configuration $F_2$) we know that
$\lambda_u^4=\lambda_y^3=2$, and  $A-E$ is a representation matrix of an
SHF$(4;n-3,m,\{2,2\})$, and there are at
most $m-2$ distinct elements in rows three and four. Similarly we have $\lceil
\frac{n-3}{m-2}\rceil \leq m-2$ and $n\leq m^2 -4m +7$.

4. $\lambda_x^3=1$, $\lambda_y^3>1$, $\lambda_u^4>1$,
$\lambda_v^4>1$.  In this case, it also holds that
$\lambda_u^4=\lambda_y^3=2$ by Lemma~\ref{1.9}(forbidden configuration $F_2$), and  $A-E$ is a representation matrix of an
SHF$(4;n-3,m,\{2,2\})$, and there are at most $m-2$ and $m-1$ distinct
elements in rows three and four respectively. Thus
we have $\lceil \frac{n-3}{m-2}\rceil \leq m-1$ and $n\leq m^2
-3m +5$.

5. $\lambda_x^3>1$, $\lambda_y^3>1$, $\lambda_u^4>1$,
$\lambda_v^4>1$. By Lemma~\ref{1.9}(forbidden configuration $F_2$) we know that
$\lambda_x^3=\lambda_y^3=\lambda_u^4=\lambda_v^4=2$, and  $A-M$ is a representation matrix of an
SHF$(4;n-4,m,\{2,2\})$, and there are at most $m-2$ distinct
elements in the last two rows. So we have $\lceil
\frac{n-4}{m-2}\rceil \leq m-2$ and $n\leq m^2 -4m +8$.

Combining the above 5 cases with the condition $m\geq 3$, we have
obtained $n\leq (m-1)^2 +1$. The proof is complete.\qed

Now we  prove that if there exists an
SHF$(4;n,4,\{2,2\})$, then $n\le 10$. To prove this conclusion, we
assume that an SHF$(4;11,4,\{2,2\})$ exists and we get a contradiction.

\begin{lemma}
\label{JG} If  $A$ is a representation matrix of an
{\rm SHF}$(4;11,4,\{2,2\})$, then $3\le \lambda_{max}\le 4$,
$d_{i,j}(x,y)\le 1$ for all admissible elements $x,y$ and
parameters $i,j$, and each row of $A$ is isomorphic to $R_1$
or $R_2$ as below.
 \begin{center}
\begin{tabular}{|c|c|c|c|c|c|c|c|c|c|c|}
\multicolumn{11}{c}{$R_1$}\\
  \hline
  $a$ & $a$ & $a$ & $a$ & $b$ & $b$ & $b$ & $\ast$  & $\ast$  & $\ast$  & $\ast$ \\
 \hline
\end{tabular}
\hspace{1cm} \begin{tabular}{|c|c|c|c|c|c|c|c|c|c|c|}
\multicolumn{11}{c}{$R_2$}\\
  \hline
  $a$ & $a$ & $a$ & $b$  & $b$  & $b$  & $c$  & $c$  & $c$  & $d$  & $d$   \\
 \hline
\end{tabular}
\end{center}
\end{lemma}

\noindent {\it Proof:} By Lemma~\ref{aabb}, $d_{i,j}(x,y)\le 1$ holds for any admissible elements $x,y$ and
parameters $i,j$. Then it is obvious that $\lambda_{max}\le 4$. Also we know that $\lambda_{max}\ge 3$
since $4\times 2=8<11$. So we have $3\le \lambda_{max}\le 4$,
and it's easy to see that each row of $A$ is isomorphic to $R_1$ or $R_2$.\qed

\begin{lemma}
\label{JGR2}If  $A$ is a representation matrix of an
{\rm SHF}$(4;11,4,\{2,2\})$, then there is at most one row of $A$ which is isomorphic to
$R_1$.
\end{lemma}

\noindent {\it Proof:} Assume, by contradiction, that $A$  has
two rows which are both isomorphic to $R_1$. Without loss of
generality, we assume both of the first two rows are isomorphic to $R_1$ and $\lambda_{a}^{1}=4$.
By Lemma~\ref{aabb}, $d_{1,2}(a,g)= 1$ holds for any element $g$ which is in the second row and in the same column with $a$. Since there are four distinct elements in total in the second row,
there exists an element $e$ such that $d_{1,2}(a,e)= 1$ and $\lambda_{e}^{2}=4$.
So we may assume the first $7$ columns of $A$ is the submatrix as below.
\begin{center}
\begin{tabular}{|c|c|c|c|c|c|c|}
  \hline
  $a$    & $a$    & $a$    & $a$ & $\ast$ & $\ast$ & $\ast$     \\
  \hline
  $\ast$ & $\ast$ & $\ast$ & $e$ & $e$    & $e$    & $e$        \\
  \hline
  $x$    & $y$    & $z$    & $t$ & $x$    & $y$    & $z$       \\
  \hline
  $u$    & $v$    & $i$    & $j$ & $a_{4,5}$ & $a_{4,6}$ & $a_{4,7}$     \\
 \hline
\end{tabular}
\end{center}

\noindent Then by Lemma~\ref{JG} we have $(a_{4,5}, a_{4,6},
a_{4,7})=(v, i, u)$ or $(i, u, v)$. We distinguish two cases.

1. $(a_{4,5}, a_{4,6},a_{4,7})=(v, i, u)$. Then we have the following submatrix.
\begin{center}
\begin{tabular}{|c|c|c|c|c|c|c|}
  \hline
  $a$    & $a$    & $a$ & $a$ & $\ast$ & $\ast$ & $\ast$   \\
  \hline
  $\ast$ & $\ast$ & $c$ & $e$ & $e$    & $e$    & $e$    \\
  \hline
  $x$    & $y$    & $z$ & $t$ & $x$    & $y$    & $z$     \\
  \hline
  $u$    & $v$    & $i$ & $j$ & $v$    & $i$    & $u$     \\
 \hline
\end{tabular}
\end{center}
Thus, the
column sets $C_1 = \{1,6\}$ and $C_2 =\{3,5\}$ are not
separable, a contradiction.

2. $(a_{4,5}, a_{4,6},a_{4,7})=(i, u, v)$. Then we also have the following submatrix.
\begin{center}
\begin{tabular}{|c|c|c|c|c|c|c|}
  \hline
  $a$    & $a$    & $a$ & $a$ & $\ast$ & $\ast$ & $\ast$    \\
  \hline
  $\ast$ & $\ast$ & $c$ & $e$ & $e$    & $e$    & $e$      \\
  \hline
  $x$    & $y$    & $z$ & $t$ & $x$    & $y$    & $z$      \\
  \hline
  $u$    & $v$    & $i$ & $j$ & $i$    & $u$    & $v$       \\
 \hline
\end{tabular}
\end{center}
Thus, the
column sets $C_1 = \{1,7\}$ and $C_2 =\{2,5\}$ are not
separable, a contradiction.\qed

\begin{lemma}
\label{JGR1} If  $A$ is a representation matrix of an
{\rm SHF}$(4;11,4,\{2,2\})$, then each row of $A$  is isomorphic to $R_2$.
\end{lemma}

\noindent {\it Proof:} By Lemma~\ref{JGR2}, we know that $A$  has
no two rows which are both isomorphic to $R_1$.
So we assume that the first row of
$A$ is isomorphic to $R_1$ and the other rows of $A$
are isomorphic to $R_2$. Without loss of generality,
we start with the following submatrix.

\begin{center}
\begin{tabular}{|c|c|c|c|c|c|c|c|c|c|c|}
  \hline
  $a$    & $a$    & $a$ & $a$ & $\ast$ & $\ast$ & $\ast$ & $\ast$  & $\ast$  & $\ast$  & $\ast$   \\
  \hline
  $b$    & $c$    & $d$ & $e$ & $b$    & $b$    & $c$    & $c$     & $d$     & $d$     & $e$   \\
  \hline
  $x$    & $y$    & $z$ & $t$ & $\ast$ & $\ast$ & $\ast$ & $\ast$  & $\ast$  & $\ast$  & $x$   \\
  \hline
  $u$    & $v$    & $i$ & $j$ & $\ast$ & $\ast$ & $\ast$ & $\ast$  & $\ast$  & $\ast$  & $\ast$   \\
 \hline
\end{tabular}
\end{center}

By Lemma~\ref{JG}, $a_{4,11}= v$ or $i$.  By Lemma~\ref{r-c}, we
only consider $a_{4,11}= v$.  By Lemma~\ref{1.9}(forbidden configuration $F_3$), considering the
column set $\{ 3, 4, 11\}$, we have
\begin{equation}\label{E5} d_{3,4}(z,v)=d_{3,4}(x,i)=0.\end{equation}

1. $\lambda_x^3=2$. Then
$\lambda_y^3=\lambda_z^3=\lambda_t^3=3$.
  By Lemma~\ref{JG} we can get the following matrix.
 \begin{center}
\begin{tabular}{|c|c|c|c|c|c|c|c|c|c|c|}
  \hline
  $a$    & $a$    & $a$ & $a$ & $\ast$ & $\ast$ & $\ast$ & $\ast$  & $\ast$  & $\ast$  & $\ast$   \\
  \hline
  $b$    & $c$    & $d$ & $e$ & $b$    & $b$    & $c$    & $c$     & $d$     & $d$     & $e$   \\
  \hline
  $x$    & $y$    & $z$ & $t$ & $z$    & $y$    & $z$    & $t$     & $y$     & $t$     & $x$   \\
  \hline
  $u$    & $v$    & $i$ & $j$ & $\ast$ & $\ast$ & $\ast$ & $\ast$  & $\ast$  & $\ast$  & $v$   \\
 \hline
\end{tabular}
\end{center}

 Since $d_{2,4}(b,u)=d_{3,4}(z,i)=1$ and (\ref{E5}), we have $a_{4,5}=j$, then $a_{4,7}=u$.
 So we have a submatrix (rows 1,2,3,4 and columns 2,4,5,7)
  which is isomorphic to the forbidden configuration $F_3$,
 a contradiction.

2. $\lambda_x^3=3$.

(i) $d_{2,3}(c,x)=1$, let $a_{3,7}=x$. Since
$d_{3,4}(x,u)=d_{3,4}(x,v)=1$,
  we have $a_{4,7}=j$ by (\ref{E5}). By Lemma~\ref{1.9}(forbidden configuration $F_3$),
  considering the column set $\{ 2, 3, 7\} $
 we have $d_{3,4}(z,j)=0$. Then $\lambda_z^3=2$£¬
 and $d_{3,4}(z,u)=1$ by (\ref{E5}). Then we have the following submatrix.
\begin{center}
\begin{tabular}{|c|c|c|c|c|c|c|c|c|c|c|}
  \hline
  $a$    & $a$    & $a$ & $a$ & $\ast$    & $\ast$ & $\ast$ & $\ast$  & $\ast$     & $\ast$  & $\ast$   \\
  \hline
  $b$    & $c$    & $d$ & $e$ & $b$       & $b$    & $c$    & $c$     & $d$        & $d$     & $e$   \\
  \hline
  $x$    & $y$    & $z$ & $t$ & $\ast$    & $\ast$ & $x$    & $z$     & $\ast$     & $\ast$  & $x$   \\
  \hline
  $u$    & $v$    & $i$ & $j$ & $\ast$    & $\ast$ & $j$    & $u$     & $\ast$     & $\ast$  & $v$   \\
 \hline
\end{tabular}
\end{center}
Thus,  the
column sets $C_1 = \{3,7\}$ and $C_2 =\{4,8\}$ are not
separable, a contradiction.

(ii) $d_{2,3}(c,x)=0$, let $a_{3,9}=x$. Since
$d_{3,4}(x,u)=d_{3,4}(x,v)=1$,
  so $a_{4,9}=j$ by (\ref{E5}).

\begin{center}
\begin{tabular}{|c|c|c|c|c|c|c|c|c|c|c|}
  \hline
  $a$    & $a$    & $a$ & $a$ & $\ast$ & $\ast$ & $\ast$ & $\ast$  & $\ast$  & $\ast$  & $\ast$   \\
  \hline
  $b$    & $c$    & $d$ & $e$ & $b$    & $b$    & $c$    & $c$     & $d$     & $d$     & $e$   \\
  \hline
  $x$    & $y$    & $z$ & $t$ & $\ast$ & $\ast$ & $\ast$ & $\ast$  & $x$     & $\ast$  & $x$   \\
  \hline
  $u$    & $v$    & $i$ & $j$ & $\ast$ & $\ast$ & $\ast$ & $\ast$  & $j$     & $\ast$  & $v$   \\
 \hline
\end{tabular}
\end{center}
\noindent Thus, the column sets $C_1=\{ 2,9\}$ and $C_2=\{ 3,11\}$
are not separable, a contradiction.\qed

\begin{lemma}
\label{333} If $A$ is a representation matrix of an
{\rm SHF}$(4;11,4,\{2,2\})$ and each row of  $A$ is isomorphic to $R_2$, then
there exists a submatrix $B$ satisfying
 $\lambda_e^2=\lambda_f^2=\lambda_g^2=3$ and
 $\lambda_x^3=\lambda_y^3=\lambda_z^3=3$.
\end{lemma}
\begin{center}
$B=$ \begin{tabular}{|c|c|c|}
  \hline
  $a$    & $a$    & $a$    \\
  \hline
  $e$    & $f$    & $g$    \\
  \hline
  $x$    & $y$    & $z$    \\
  \hline
\end{tabular}
\end{center}
\noindent {\it Proof:} Suppose that  $A$ does not contain a submatrix  isomorphic to $B$.
Then we  show that $A$ is not a representation matrix of an SHF$(4;11,4,\{2,2\})$.
Since each row of $A$ is isomorphic to
$R_2$, we have $\lambda_d^1=\lambda_h^2=\lambda_t^3=\lambda_j^4=2$.
By Lemma~\ref{r-c},  we have the following submatrix, where $C$,
$D$ and $E$ are $3\times 3$ matrices,
and $F$ is a $3\times 2$ matrix.
\begin{center}
\begin{tabular}{|c|c|c|c|c|c|c|c|c|c|c|}
  \hline
  $a$    & $a$  & $a$ & $b$ & $b$  & $b$  & $c$ & $c$  & $c$  & $d$  & $d$   \\
  \hline
  \multicolumn{3}{|c|}{C} & \multicolumn{3}{|c|}{D}   & \multicolumn{3}{|c|}{E}     &  \multicolumn{2}{|c|}{F}   \\
  \hline
\end{tabular}
\end{center}
If there is not element in $\{h,t,j\}$ satisfying
$d_{1,k}(a,g)=1$, then the submatrix (row set $\{1,2,3\}$ and column set $\{1,2,3\}$)  is isomorphic to $B$;
If there exists  exactly one element $g$ in $\{h,t,j\}$ satisfying
$d_{1,k}(a,g)=1$, then the submatrix (row set $\{1,2,3,4\}\backslash \{k\}$ and column set $\{1,2,3\}$)  is isomorphic to $B$.
  So the matrix $C$ contains at least two elements in $\{h,t,j\}$.
   Similarly, the matrices  $D$ and $E$  also contain at least two elements in $\{h,t,j\}$, respectively.
 Since  $\lambda_h^2=\lambda_t^3=\lambda_j^4=2$,
 each of matrices of $C, D$ and $E$ contains two elements of $\{h,t,j\}$.
  So the matrix $F$ does not contain any  one element of $\{h,t,j\}$.
Thus,  for any two elements $g, g_1$ in $\{d,h,t,j\}$,
$d_{k_1,k_2}(g,g_1)=0, \{k_1, k_2\}\subset\{1,2,3,4\} $.
 By Lemma~\ref{r-c},  we have the following submatrix.
  \begin{center}
\begin{tabular}{|c|c|c|c|c|c|c|c|c|c|c|}
  \hline
  $d$    & $d$  &  $\ast$   &  $\ast$  &   $\ast$ &   $\ast$  &  $\ast$  &   $\ast$  &   $a$  & $b$  & $c$     \\
  \hline
    $\ast$   &   $\ast$   & $h$ & $h$ & $\ast$ & $\ast$  &  $\ast$ &  $\ast$ & $e$    & $f$  & $h$   \\
  \hline
     $x$ & $y$  & $x$ & $z$ & $t$ & $t$&  $\ast$&  $\ast$ & $x$    & $y$  & $z$\\
  \hline
     $v$ & $\ast$& $i$ &  $\ast$ & $\ast$&   $\ast$ &$j$ & $j$ & $u$    & $v$  & $i$  \\
  \hline
\end{tabular}
\end{center}
Since $ a_{4,2}$ and $a_{4,4}$ are in the set $\{ u,v,i\}$,
we have $(a_{4,2},a_{4,4})\in\{(u,u ),(i,u ),(u,v)\}$.
 If $(a_{4,2},a_{4,4})= (u,u )$, then the column sets $C_1=\{ 1,4\}$ and $C_2=\{ 2,3\}$
are not separable, a contradiction;
If $(a_{4,2},a_{4,4})= (i,u )$,
then $(a_{1,3},a_{1,4})= (b,b )$ by Lemma \ref{JG}, a contradiction;
If $(a_{4,2},a_{4,4})= (u,v )$,
then $(a_{2,1},a_{2,2})= (e,e )$ by Lemma \ref{JG}, a contradiction.\qed

\begin{lemma}
\label{JGR0}
There is no {\rm SHF}$(4;11,4,\{2,2\})$ with all $4$ rows isomorphic to $R_2$.
\end{lemma}
\noindent {\it Proof:} Let $A$ is a representation matrix of an
SHF$(4;11,4,\{2,2\})$. Assume that every row of $A$ is
 isomorphic to $R_2$. By Lemma~\ref{333}, we suppose the first three rows and columns is a
 submatrix which is isomorphic to $B$ satisfying
 $\lambda_e^2=\lambda_f^2=\lambda_g^2=3$ and
 $\lambda_x^3=\lambda_y^3=\lambda_z^3=3$. Suppose the fourth
 elements
 in rows two and three are $h$ and $t$ respectively. Then
 $\lambda_h^2=\lambda_t^3=2$. We distinguish two cases.

1. $d_{2,3}(h,t)=0$. Without loss of generality, let
$d_{2,3}(e,t)=d_{2,3}(f,t)=1$.

(i) $a_{3,5}=a_{3,7}$. Since $d_{2,3}(e,x)=d_{2,3}(f,y)=1$, we have
$a_{3,5}=a_{3,7}=z$. So we have the following submatrix.
\begin{center}
\begin{tabular}{|c|c|c|c|c|c|c|c|c|c|c|}
  \hline
  $a$    & $a$  & $a$ & $\ast$ & $\ast$ & $\ast$ & $\ast$ & $\ast$  & $\ast$  & $\ast$  & $\ast$   \\
  \hline
  $e$    & $f$  & $g$ & $e$    & $e$    & $f$    & $f$    & $g$     & $g$     & $h$     & $h$   \\
  \hline
  $x$    & $y$  & $z$ & $t$    & $z$    & $t$    & $z$    & $x$     & $y$     & $x$     & $y$   \\
  \hline
  $u$    & $v$  & $i$ & $\ast$ & $\ast$ & $\ast$ & $\ast$ & $\ast$  & $\ast$  & $\ast$  & $\ast$  \\
 \hline
\end{tabular}
\end{center}
By Lemma~\ref{JG}, we have $i\notin \{a_{4,5}, a_{4,7}, a_{4,8}, a_{4,9}\}$.
By Lemma~\ref{1.9}(forbidden configuration $F_3$), we have  $i\notin\{a_{4,4}, a_{4,6}, a_{4,10}, a_{4,11}\}$,
otherwise, we have 4 submatrices (row set $\{1, 2, 3, 4\}$,
and column sets $\{2, 3, 4, 6 \}$, $\{1, 3, 4, 6 \}$, $\{2, 3, 10, 11 \}$, $\{1, 3, 10, 11 \}$ respectively) which are all isomorphic to  $F_3$.
So we have $\lambda_i^4=1$, a contradiction.

(ii) $a_{3,5}\not=a_{3,7}$. We distinguish three cases,
$(a_{3,5},a_{3,7})=(y,x),(y,z)$, and $(z,x)$.
 If $(a_{3,5},a_{3,7})=(y,x)$, then $(a_{3,10},a_{3,11})=(z,z)$, a contradiction.
 $(a_{3,5},a_{3,7})=(y,z)$ and $(a_{3,5},a_{3,7})=(z,x)$ are isomorphic.
  So let$(a_{3,5},a_{3,7})=(z,x)$.
\begin{center}
\begin{tabular}{|c|c|c|c|c|c|c|c|c|c|c|}
  \hline
  $a$    & $a$  & $a$ & $\ast$ & $\ast$ & $\ast$ & $\ast$ & $\ast$  & $\ast$  & $\ast$  & $\ast$   \\
  \hline
  $e$    & $f$  & $g$ & $e$    & $e$    & $f$    & $f$    & $g$     & $g$     & $h$     & $h$   \\
  \hline
  $x$    & $y$  & $z$ & $t$    & $z$    & $t$    & $x$    & $x$     & $y$     & $y$     & $z$   \\
  \hline
  $u$    & $v$  & $i$ & $\ast$ & $\ast$ & $\ast$ & $\ast$ & $\ast$  & $\ast$  & $\ast$  & $\ast$  \\
 \hline
\end{tabular}
\end{center}
By Lemmas~\ref{1.9} and \ref{JG}, the possible elements of each position in the last row
 are listed as below.
\begin{center}
\begin{tabular}{|c|c|c|c|c|c|c|c|c|}
  \hline
   $ a_{4,k}$ & $a_{4,4}$    & $a_{4,5}$  &  $a_{4,6}$ &$a_{4,7}$ & $a_{4,8}$ & $a_{4,9}$  & $a_{4,10}$  & $a_{4,11}$    \\
  \hline
  possible elements   & $v,j$       & $v,j$     & $u,j$     & $i,j$   & $v,j$    & $u,j$     & $i,j$       & $u,j$             \\
 \hline
\end{tabular}
\end{center}

If $a_{4,4}=v$, by Lemma~\ref{JG}, we have $a_{4,5}=j$, then
$a_{4,11}=u$. By Lemma~\ref{1.9}(forbidden configuration $F_3$), considering the column set $\{
1, 2, 5\}$,
 we have $d_{3,4}(y,j)=0$, then $a_{4,9}=u$. So
 $a_{4,6}=j$ by $\lambda_{u}^4=3$.
 By Lemma~\ref{1.9}(forbidden configuration $F_3$), considering the column set $\{ 2, 3, 6\}$,
 we have $d_{3,4}(z,j)=0$, a contradiction.

If $a_{4,4}=j$, by Lemma~\ref{JG} we have $a_{4,5}=v$ and
$a_{4,6}=u$. By Lemma~\ref{1.9}(forbidden configuration $F_3$), considering the column sets $\{
1, 2, 4\}$ and
 $\{ 1, 3, 4\}$, we have $d_{3,4}(y,j)=d_{34}(z,j)=0$, so $a_{4,9}=a_{4,11}=u$.
Thus, $\lambda_{u}^4=4$, a contradiction.

2.  $d_{2,3}(h,t)=1$. Without loss of generality, let
$d_{2,3}(g,t)=1$.
\begin{center}
\begin{tabular}{|c|c|c|c|c|c|c|c|c|c|c|}
  \hline
  $a$    & $a$  & $a$ & $\ast$ & $\ast$ & $\ast$ & $\ast$ & $\ast$  & $\ast$  & $\ast$  & $\ast$   \\
  \hline
  $e$    & $f$  & $g$ & $e$    & $e$    & $f$    & $f$    & $g$     & $g$     & $h$     & $h$   \\
  \hline
  $x$    & $y$  & $z$ & $y$    & $z$    & $x$    & $z$    & $x$     & $t$     & $t$     & $y$   \\
  \hline
  $u$    & $v$  & $i$ & $\ast$ & $\ast$ & $\ast$ & $\ast$ & $\ast$  & $\ast$  & $\ast$  & $\ast$  \\
 \hline
\end{tabular}
\end{center}
Similarly, we have the following table.
\begin{center}
\begin{tabular}{|c|c|c|c|c|c|c|c|c|}
  \hline
  $ a_{4,k}$         & $a_{4,4}$    & $a_{4,5}$  &  $a_{4,6}$ &$a_{4,7}$ & $a_{4,8}$ & $a_{4,9}$  & $a_{4,10}$  & $a_{4,11}$    \\
  \hline
possible elements   & $i,j$       & $v,j$     & $i,j$     & $u,j$   & $v,j$    & $u,v,j$   & $j$         & $u,j$             \\
 \hline
\end{tabular}
\end{center}
By Lemma~\ref{JG}, we have $a_{4,9}=u$ and $a_{4,11}=u$.
Then $\lambda_u^4=3$, and $a_{4,7}=j$.
By Lemma~\ref{JG}, we have $a_{4,5}=v$ and $a_{4,6}=i$.

\begin{center}
\begin{tabular}{|c|c|c|c|c|c|c|c|c|c|c|}
  \hline
  $a$    & $a$  & $a$ & $\ast$ & $\ast$ & $\ast$ & $\ast$ & $\ast$  & $\ast$  & $\ast$  & $\ast$   \\
  \hline
  $e$    & $f$  & $g$ & $e$    & $e$    & $f$    & $f$    & $g$     & $g$     & $h$     & $h$   \\
  \hline
  $x$    & $y$  & $z$ & $y$    & $z$    & $x$    & $z$    & $x$     & $t$     & $t$     & $y$   \\
  \hline
  $u$    & $v$  & $i$ &$\ast$  & $v$    & $i$    & $j$    & $\ast$  & $u$     & $j$    & $u$  \\
 \hline
\end{tabular}
\end{center}
Now, we have a submatrix (rows 1, 2, 3, 4
and columns 2, 3, 10, 11) which is isomorphic to  $F_3$, a contradiction. \qed

\begin{theorem}
\label{O10}
There exists an optimal {\rm SHF}$(4;10,4,\{2,2\})$.
\end{theorem}
\noindent {\it Proof:} It's obvious that Lemma \ref{JGR0} contradicts with Lemma \ref{JGR1}.
So if there is an  SHF$(4;n,4,\{2,2\})$, then $n \leq 10$.
Next, we give a representation matrix of an SHF$(4;10,4,\{2,2\})$ in the following.

\begin{center}
\begin{tabular}{|c|c|c|c|c|c|c|c|c|c|}
  \hline
  1   & 1  & 1 & 2 & 2   & 2  & 3  & 3  & 3  & 4    \\
  \hline
  1   & 2  & 3 & 1 & 2   & 3  & 1  & 2  & 3  & 4    \\
  \hline
  1   & 2  & 3 & 2 & 3   & 1  & 3  & 1  & 2  & 4    \\
  \hline
  1   & 2  & 3 &3  & 1   & 2  & 2  & 3  & 1  & 4    \\
  \hline
\end{tabular}
\end{center}

Thus, there exists an optimal {\rm SHF}$(4;10,4,\{2,2\})$. \qed

\begin{theorem}
\label{SHFM} If there exists an {\rm SHF}$(4;n,m,\{2,2\})$ with $m \geq
4$, then $n\le (m-1)^2+1$.
\end{theorem}

\noindent {\it Proof:} If $m=4$, the conclusion follows by Theorem~\ref{O10}.
Now assume that $m>4$.
 Let $A$ be a representation matrix of an SHF$(4;n,m,\{2,2\})$.
Now, we consider the following two cases.

1. There is a pair of elements $x$ and $y$ in the $i$-th row and the  $j$-th row respectively such that $d_{i,j}(x,y)>1$.
 Then we have $n\leq (m-1)^2+1$ by Lemma~\ref{aabb}.

2. $d_{i,j}(x,y)\le 1$ for any admissible elements $x,y$ and
parameters $i,j$. Then we have $\lambda_{max}\leq m$.
Assume, for a contradiction that $n= (m-1)^2+2$.
By the pigeonhole principle, there is an element $t_i$ such that
 $\lambda_{t_i}^i\ge \lceil \frac{n}{m}\rceil= m-1$ for $1\leq i\le 4$.

(i) $\lambda_{max}= m$.
Assume that there are  two elements $a$ and $k$ in different rows (without loss of generality, in the first two rows)
such that $\lambda_{a}^1=\lambda_{k}^2=m$.
Then there is a submatrix of $A$ as below.
\begin{center}
\begin{tabular}{|c|c|c|c|c|c|c|c|c|}
  \hline
  $a$   & $a$     & $\cdots$ & $a$        & $a$    & $\ast$   & $\ast$   & $\cdots$ & $\ast$      \\
  \hline
  $\ast$& $\ast$  & $\cdots$ & $\ast$     & $k$    & $k$      & $k$      & $\cdots$ & $k$     \\
  \hline
  $x_1$ & $x_2$& $\cdots$    & $x_{m-1}$  & $x_m$  & $y_1$    & $y_2$    &$\cdots$  & $y_{m-1}$     \\
  \hline
  $u_1$ & $u_2$& $\cdots$    & $u_{m-1}$  & $u_m$  & $v_1$    & $v_2$    & $\cdots$ & $v_{m-1}$      \\
  \hline
\end{tabular}
\end{center}
By Lemma~\ref{1.9}(forbidden configuration $F_3$), $d_{3,4}(x_i,v_j)=0$ and $d_{4,3}(u_i,y_j)=0$ for any $u_i\not=v_j$, $x_i\not=y_j$, $1\le i,j\le m-1$.
So, there are  at least $(m-1)\times (m-3)$ distinct pairs of elements
  $s$ and $t$ such that $d_{3,4}(s,t)=0$.
 Thus,  $(m-1)\times (m-3) + n \le m^2$,
 ie. $(m-1)\times (m-3) + (m-1)^2+2 \le m^2$,
 so we have $m\le 4$. It contradicts $m>4$.

So we may assume that there is exactly one row (without loss of generality, the first row) containing an element $a$
such that $\lambda_{a}^1=m$.
\begin{center}
\begin{tabular}{|c|c|c|c|c|c|c|c|c|}
  \hline
  $a$   & $a$     & $\cdots$ & $a$       & $a$      & $\ast$   & $\ast$    & $\cdots$  & $\ast$      \\
  \hline
  $\ast$& $\ast$  & $\cdots$ & $\ast$    & $k$      & $k$      & $k$       & $\cdots$  & $k$     \\
  \hline
  $x_1$ & $x_2$   & $\cdots$ & $x_{m-1}$ & $x_m$    & $y_1$    & $y_2$     &$\cdots$   & $y_{m-2}$     \\
  \hline
  $u_1$ & $u_2$   & $\cdots$ & $u_{m-1}$ & $u_m$    & $v_1$    & $v_2$     & $\cdots$  & $v_{m-2}$      \\
  \hline
\end{tabular}
\end{center}
Similarly, we can obtain $(m-2)\times (m-3)$ distinct pairs of
elements $s$ and $t$
  such that $d_{3,4}(s,t)=0$. Since $\lambda_{x_m}^3\le m-1$ and $\lambda_{u_m}^4\le
  m-1$, we know that there are two elements $w$ in row three and $z$
  in row four such that $d_{3,4}(x_m,z)=d_{3,4}(w,u_m)=0$.
 Thus, $(m-2)\times (m-3)+2 + n \le m^2$
  contradicts $m>4$.

(ii) $\lambda_{max}= m-1$. Then  there exist  two elements $a$ and
$k$ in the first two rows such that
$\lambda_{a}^1=\lambda_{k}^2=m-1$. Suppose there is a submatrix of
$A$ as blew.
 \begin{center}
\begin{tabular}{|c|c|c|c|c|c|c|c|c|}
  \hline
  $a$     & $a$      & $\cdots$   & $a$       & $a$       & $\ast$   & $\ast$    & $\cdots$ & $\ast$      \\
  \hline
 $\ast$   & $\ast$   & $\cdots$   & $\ast$    & $k$       & $k$      & $k$       & $\cdots$ & $k$     \\
  \hline
  $x_1$   & $x_2$    & $\cdots$   & $x_{m-2}$ & $x_{m-1}$ & $y_1$    & $y_2$     &$\cdots$  & $y_{m-2}$     \\
  \hline
  $u_1$   & $u_2$    & $\cdots$   & $u_{m-2}$ & $u_{m-1}$ & $v_1$    & $v_2$     & $\cdots$ & $v_{m-2}$      \\
  \hline
\end{tabular}
\end{center}
Similarly, there are at least $(m-2)\times (m-4)+ 2$ distinct
pairs of elements $s$ and $t$
  such that $d_{3,4}(s,t)=0$.
 Thus, $(m-2)\times (m-4)+2 + n \le m^2$.
 So we have $m\le 5$.
If $m>5$, we have a contradiction.
Now we assume that $m=5$.
Then we have the following submatrix of $A$.
\begin{center}
\begin{tabular}{|c|c|c|c|c|c|c|}
  \hline
  $a$     & $a$         & $a$       & $a$       & $\ast$   & $\ast$    & $\ast$      \\
  \hline
 $\ast$   & $\ast$      & $\ast$    & $k$       & $k$      & $k$       & $k$     \\
  \hline
  $x_1$   & $x_2$       & $x_{3}$   & $x_{4}$   & $y_1$    & $y_2$     & $y_{3}$     \\
  \hline
  $u_1$   & $u_2$       & $u_{3}$   & $u_{4}$   & $v_1$    & $v_2$     & $v_{3}$      \\
  \hline
\end{tabular}
\end{center}
Let $X_1=\{x_1, x_2, x_3\}$,
$Y_1=\{y_1, y_2,  y_3\}$,
$U_1=\{u_1, u_2,  u_3\}$ and
$V_1=\{v_1, v_2,  v_3\}$. Since $d_{i,j}(x,y)\le 1$, we have $|\{x_1,x_2,x_3,x_4\}|=|\{x_4,y_1,y_2,y_3\}|=4$.
Then by $m=5$ we have $2\le \mid X_1\cap Y_1 \mid \le 3$. Similarly, we can get $2\le \mid U_1\cap V_1 \mid \le 3$.
Now we continue to distinguish the following 2 cases.

(a) $\mid X_1\cap Y_1 \mid = 3$ and $\mid U_1\cap V_1 \mid =3$.
Thus, $X_1=Y_1$ and $U_1= V_1$.
By Lemma \ref{r-c} we only need to consider the following two submatrices.
\begin{center}
\begin{tabular}{|c|c|c|c|c|c|c|}
  \hline
  $a$     & $a$         & $a$       & $a$       & $\ast$   & $\ast$    & $\ast$      \\
  \hline
 $\ast$   & $\ast$      & $\ast$    & $k$       & $k$      & $k$       & $k$     \\
  \hline
  $x_1$   & $x_2$       & $x_{3}$   & $x_{4}$   & $x_1$    & $x_2$     & $x_{3}$     \\
  \hline
  $u_1$   & $u_2$       & $u_{3}$   & $u_{4}$   & $u_2$    & $u_3$     & $u_{1}$      \\
  \hline
\end{tabular}
\hspace{1cm}
\begin{tabular}{|c|c|c|c|c|c|c|}
  \hline
  $a$     & $a$         & $a$       & $a$       & $\ast$   & $\ast$    & $\ast$      \\
  \hline
 $\ast$   & $\ast$      & $\ast$    & $k$       & $k$      & $k$       & $k$     \\
  \hline
  $x_1$   & $x_2$       & $x_{3}$   & $x_{4}$   & $x_1$    & $x_2$     & $x_{3}$     \\
  \hline
  $u_1$   & $u_2$       & $u_{3}$   & $u_{4}$   & $u_3$    & $u_1$     & $u_{2}$      \\
  \hline
\end{tabular}
\end{center}
 It is easy to check that in the left submatrix
column sets $C_1 = \{1,6\}$ and $C_2 =\{3,5\}$ are not
separable, and  in the right submatrix the column sets $C_1 = \{1,7\}$ and
$C_2 =\{2,5\}$ are not separable, a contradiction.

(b) $\mid X_1\cap Y_1 \mid = 2$ and $\mid U_1\cap V_1 \mid \le 3$.
By Lemma \ref{r-c}, we may assume that $y_1=x_1, y_2=x_2$ and $y_3=x_5$.
So we have the following submatrix.
\begin{center}
\begin{tabular}{|c|c|c|c|c|c|c|}
  \hline
  $a$     & $a$         & $a$       & $a$       & $\ast$   & $\ast$    & $\ast$      \\
  \hline
 $\ast$   & $\ast$      & $\ast$    & $k$       & $k$      & $k$       & $k$     \\
  \hline
  $x_1$   & $x_2$       & $x_{3}$   & $x_{4}$   & $x_1$    & $x_2$     & $x_5$     \\
  \hline
  $u_1$   & $u_2$       & $u_{3}$   & $u_{4}$   & $v_1$    & $v_2$     & $v_3$      \\
  \hline
\end{tabular}
\end{center}

For each $i=1,2$, since $|V_1\setminus \{v_i,u_i\}|\ge 1$, we know that there is  at least one element
  $t_i\in V_1\setminus \{v_i,u_i\}$ such that $d_{3,4}(x_i,t_i)=0$  by Lemma~\ref{1.9}(forbidden configuration $F_3$).

Similarly, since $|V_1\setminus \{u_3\}|\ge 2$ and $|U_1\setminus \{v_3\}|\ge 2$ we know that there are  at least two elements
  $t_3,t_4\in V_1\setminus \{u_3\}$ and $t_5,t_6\in U_1\setminus \{v_3\}$ such that $d_{3,4}(x_3,t_3)=0$, $d_{3,4}(x_3,t_4)=0$ and
  $d_{3,4}(x_5,t_5)=0$, $d_{3,4}(x_5,t_6)=0$ respectively.

Since $m=5, \lambda_{x_4}^3\le 4$, and $\lambda_{u_4}^4\le 4$,
we know that there are two elements $w$ in row three and $z$
  in row four such that $w\not=x_4$, $z\not=u_4$ and $d_{3,4}(x_4,z)=d_{3,4}(w,u_4)=0$.

  Since $x_4\not\in X_1\cup Y_1$ and $u_4\not\in U_1\cup V_1$
we have obtained at least $8$ distinct pairs of elements
  $(s,t)$ such that $d_{3,4}(s,t)=0$.
So we have   $8+n=8 + 18=26>25$,
 a contradiction.

The proof is complete.\qed

\noindent {\bf Remark 2 :} It's easy to see that the bound in Theorem~\ref{SHFM} is tight for $m=4$ by Theorem~\ref{O10}.

Now we are in the position to prove the main result of this
section.

\begin{theorem}
\label{WW}
If there exists an {\rm SHF}$(2w;n,m,\{w,w\})$
with $m \geq 2w \geq 4$, then $n\le (m-1)^2+1$.
\end{theorem}
\noindent {\it Proof:} We use induction on $w$ to prove the theorem.

1. By Theorem~\ref{SHFM}, for $w=2$ this conclusion holds.

2. Assume  the conclusion holds for $w=k-1$, $k\ge 3$.
Then, we consider the case $w=k$. Let $m\ge 2k\ge 6$.
Assume, for a contradiction, that $A$ is  a representation matrix of an SHF$(2k;(m-1)^2+2,m,\{k,k\})$, and
 $\C$ denote the set of all columns of $A$. By removing the first two rows of $A$,
we obtain a $(2k-2)\times (m^2-2m+3)$ submatrix $B$. By
inductive hypothesis, $B$ is not an SHF$(2k-2;(m-1)^2+2,m,\{k-1,k-1\})$.
Thus,  there are two disjoint subsets of columns $C_1$ and
$C_2$ of $B$ with $|C_1|= |C_2|=k -1$ which are not
separable. Now, we consider the same column sets $C_1$ and $C_2$
in $A$.
 Let $D=\C \backslash (C_1\cup C_2)$, then $|D|=m^2-2m-2k+5\geq m^2-3m+5>m$.
Let $Y_i,U_i$ and $V_i$ $(i=1,2)$
be the element sets in which each element appears in the $i$-th row
of these columns in $D, C_1$ and $C_2$ respectively.

If there exist two elements $x\in Y_1$ and $y\in Y_2$ such that $d_{1,2}(x,y)\ge 2$,
then we have two distinct columns $l_1$ and $l_2$ agreeing in the first two rows.
So $C_1\cup\{l_1\}$ and $C_2\cup\{l_2\}$ are not separated in $A$, a contradiction.
Now, assume that $d_{1,2}(x,y)\le 1$ for all $x\in Y_1$ and $y\in Y_2$.
If $\{a_{1,j_1}: j_1\in C_1\}\cap \{a_{1,j_2}: j_2\in C_2\}\neq \emptyset$,
then we have two distinct columns $l_1$ and $l_2$ agreeing in the second row.
So $C_1\cup\{l_1\}$ and $C_2\cup\{l_2\}$ are not separated in $A$, a contradiction.
By  Lemma \ref{r-c}, we only need to consider the case that
$\{a_{i,j_1}: j_1\in C_1\}\cap \{a_{i,j_2}: j_2\in C_2\}=\emptyset, 1\le i\le 2$,
and $d_{1,2}(x,y)\le 1$ for all $x\in Y_1$ and $y\in Y_2$.
Suppose that $|Y_1|\ge |Y_2|$, we distinguish  the following four cases.

(i) $|Y_1|\le m-1$.  Since $\lceil\frac{n-2(k-1)}{m-1}\rceil= m-1$,
by the pigeonhole principle, there exists an element $x\in Y_1$ such that $\lambda_{x}^{1}\ge m-1$ in $D$.
Since $d_{1,2}(x,y)\le 1$ for $x\in Y_1$ and $y\in Y_2$, we have two columns $l_1$ and
$l_2$ in $D$ such that $ a_{1,l_1}=a_{1,l_2}=x  $ and  $a_{2,l_1} \in U_2 \cup V_2$.
If  $a_{2,l_1} \in U_2$, then $C_1\cup\{l_2\}$ is not separated from $C_2\cup\{l_1\}$,  a contradiction.
Similarly, we have a contradiction for $ a_{2,l_1} \in V_2$.

(ii) $|Y_1|=m$ and $|Y_2|< m-1$.  Since $\lceil\frac{n-2(k-1)}{m-2}\rceil= m$,
by the pigeonhole principle, there exists an element $x\in Y_2$ such that $\lambda_{x}^{2}\ge m$ in $D$.
Then we have two columns $l_1$ and
$l_2$ in $D$ such that $ a_{2,l_1}=a_{2,l_2}=x  $ and $ a_{1,l_1} \in U_1$.
So  $C_1\cup\{l_2\}$ is not separated from $C_2\cup\{l_1\}$,  a contradiction.

(iii) $|Y_1|=m$ and $|Y_2|= m-1$. Since $\lceil\frac{n-2(k-1)}{m-1}\rceil= m-1$,
by the pigeonhole principle, there exists an element $x\in Y_2$ such that $\lambda_{x}^{2}\ge m-1$ in $D$.
 We may have a contradiction by using the similar method in case (i).

(iv) $|Y_1|=|Y_2|=m$. For any column $k_i\in C_i$ $(i=1,2)$, we
have  $a_{1,k_1} \in Y_1 $, $a_{1,k_2}\in Y_1$, $a_{2,k_1} \in
Y_2$ and $a_{2,k_2}\in Y_2$. If there exist distinct columns $l_1$
and $l_2$ in $D$, such that $a_{1,k_2}=a_{1,l_1}$  and
$a_{2,k_1}=a_{2,l_2}$, we  have that $C_1\cup\{l_1\}$ is not
separated from $C_2\cup\{l_2\}$, a contradiction. Otherwise, there  exists the unique
column $l$ such that $a_{1,k_2}=a_{1,l}$, or
$a_{2,k_1}=a_{2,l}$, so $a_{1,k_2}=a_{1,l}$ and
$a_{2,k_1}=a_{2,l}$.
Similarly, we have a unique column  $l^\prime $
such that $a_{1,k_1}=a_{1,l^\prime}$ and $a_{2,k_2}=a_{2,l^\prime}$.
Then there are $m-2$ distinct elements in the first row of column set $
D\backslash \{l,l^\prime\}$
 and $m-2$ distinct elements in the second row of column set $ D\backslash \{l,l^\prime\}$.
Since $\lceil\frac{n-2(k-1)-2}{m-2}\rceil=m-1> m-2$, by the pigeon
hole principle, there exist two elements $x\in Y_1$ and $y\in Y_2$ such that $d_{1,2}(x,y)> 1$,
a contradiction.

The proof is complete.\qed

\noindent {\bf Remark 3 :}
The best upper bound for an SHF$(2w; n,m,\{w,w\})$ with $m \geq 2w \geq 4$ is $n < m^2$ \cite{BT}.
We improve this bound from $n< m^2$ to $n\le (m-1)^2+1$.

\section{An improved bound for SHF$(\sum_{i=1}^{t}w_i;n,m,\{w_1,w_2,\ldots,w_t\})$}

In this section, we shall give a new bound for an
SHF$(w_1+w_2;n,m,\{w_1,w_2\})$ with $w_2>w_1\ge 2$ and $m\ge w_1+w_2 $.
Then we obtain a new bound for SHF$(\sum_{i=1}^{t}w_i;n,m,\{w_1,w_2,\ldots,w_t\})$.

\begin{lemma}
\label{4.3aabb} Suppose $A$ is a representation matrix of an
{\rm SHF}$(2+w;n,m,\{2,w\})$ with $m \geq 2+w $. If there is a pair of
elements $a$ and $b$ such that $d_{i,j}(a,b)\ge 2$, then $ n<
m^2-m$.
\end{lemma}
\noindent {\it Proof:} Suppose, for a contradiction, that $A$ is a representation matrix of an
SHF$(2+w;m^2-m,m,\{2,w\})$ with $m \geq 2+w $, and
$d_{12}(a,b)\ge 2$.
Let $\C$ denote the set of columns of $A$, and
$ \B =\{(i,j): \lambda^i_{a_{i,j}}=1, 3\le i\le w+2, 1\le j\le 2\}$.
We distinguish  the following $4$ cases.

1. $|\B|=0$.
  Then there are two columns $l_i$  and $l_i^{\prime}$ of $A$ such that $a_{i,1}=a_{i,l_i}$ and $a_{i,2}=a_{i,l_i^{\prime}}$
 for each $ 3\le i\le w+2$. Let $C_1= \{2\}\cup\{l_i: 3\le i\le w+2 \}$ and $C_2= \{1\}\cup\{l_i^{\prime}: 3\le i\le w+2 \}$.
If there is a column $h_1\not\in C_1$ or $h_2\not\in C_1$ satisfying
$a_{w+1,2}=a_{w+1,h_1}$ or $a_{w+2,2}=a_{w+2,h_2}$ respectively,
then $C_1\backslash \{l_{w+j}\}$ is not separated from $\{1, h_j\}$ for $j=1$ or $2$, a contradiction.
Thus, for any column $l$ satisfying   $a_{w+1,2}=a_{w+2,l}$ or $a_{w+2,2}=a_{w+2,l}$
we have   $l\in C_1$.
Similarly, if   $a_{w+1,1}=a_{w+2,l^{\prime}}$ or $a_{w+2,1}=a_{w+2,l^{\prime}}$,
we have   $l^{\prime}\in C_2$.
Let $\C_1=\C\backslash (C_1\cup C_2)$.
By the pigeonhole principle, there is an
element $t$ in row $w+2$ such that $\lambda_t^{w+2}\ge \lceil \frac{n-2w-2}{m-2}\rceil>m-2$.
So there exist two columns $l_1^{\prime}$ and $l_2^{\prime}$ of ${\cal C}_1$ agreeing in the last two rows.
Then  $\{1, l_1^{\prime}\}$ and $C_1\cup \{l_2^{\prime}\}\backslash \{l_{w+1},l_{w+2}\}$  are not separated, a contradiction.

2. $|\B|=1$. By Lemma~\ref{r-c},
we may assume $\lambda_{a_{w+2,1}}^{w+2}= 1$,
$\lambda_{a_{w+2,2}}^{w+2}\ge 2$
and $\lambda^i_{a_{i,j}}\ge 2, 3\le i\le w+1, 1\le j\le 2$.
  Then there exist a column $l_i$ of $A$ such that $a_{i,1}=a_{i,l_i}$
 for each $ 3\le i\le w+1$. Let $C_1= \{2\}\cup \{l_i: 3\le i\le w+1 \}$.
If there exists a column $l$ such that
$a_{w+2,2}=a_{w+2,l}$ with $l\notin C_1$,
then let $C_2=\{1, l\}$, so
we have $C_1$ is not separated from $C_2$, a contradiction.
Thus, for any column $l$ satisfying   $a_{w+2,2}=a_{w+2,l}$,
we have  $l\in C_1$. Let $\C_1=\C\backslash ( C_1\cup\{1\} )$.
By the pigeonhole principle, there is an
element $t$ in row $w+2$ such that $\lambda_t^{w+2}\ge \lceil \frac{n-w-1}{m-2}\rceil>m$.
So there exist two columns $l_1$ and $l_2$ of ${\cal C}_1$ agreeing in the last two rows.
Then  $\{1, l_1\}$ and $C_1\cup \{l_2\}\backslash \{l_{w+1}\}$  are not separated, a contradiction.

3. $|\B|=2$. Let $\lambda_{x}^{i_1}=\lambda_{y}^{i_2}=1$.

(i) $i_1=i_2$. By Lemma~\ref{r-c}, we may assume $i_1=i_2=w+2$.  Then there exist a column $l_i$ of $A$ such that $a_{i,1}=a_{i,l_i}$
 for each $ 3\le i\le w$. Let $C_1= \{2\}\cup \{l_i: 3\le i\le w \}$.
 Let $\C_1=\C\backslash ( C_1\cup\{1\} )$.
By the pigeonhole principle, there is an
element $t$ in row $w+2$ such that $\lambda_t^{w+2}\ge \lceil \frac{n-w}{m-2}\rceil> m$.
So there exist two columns $l_1$ and $l_2$ of ${\cal C}_1$ agreeing in the last two rows.
Then  $\{1, l_1\}$ and $C_1\cup \{l_2\}$  are not separated, a contradiction.

(ii) $i_1\neq i_2$. By Lemma~\ref{r-c}, we may assume $i_1=w+1$ and $i_2=w+2$.  Then there exist a column $l_i$ of $A$ such that $a_{i,1}=a_{i,l_i}$
 for each $ 3\le i\le w$. Let $C_1= \{2\}\cup \{l_i: 3\le i\le w \}$.
 Let $\C_1=\C\backslash ( C_1\cup\{1\} )$.
By the pigeonhole principle, there is an
element $t$ in row $w+2$ such that $\lambda_t^{w+2}\ge \lceil \frac{n-w}{m-1}\rceil> m-1$.
So we have two columns $l_1$ and $l_2$ agreeing in the last two rows, and
 $\{1, l_1\}$ and $C_1\cup \{l_2\}$  are not separable, a contradiction.

4. $|\B|\ge 3$.
Let $3\le k\le w$. By Lemma~\ref{r-c}, we may assume that $\lambda_{a_{i,j}}^{i}>1$ when $3\le i\le k$
 and $1\in\{\lambda_{a_{i,1}}^{i},\lambda_{a_{i,2}}^{i}\}$  when $k+1\le i\le w+2$.
Then there exist a column $l_i$ of $A$ such that $a_{i,1}=a_{i,l_i}$
 for each $3\le i\le k$. Let $C_1= \{2\}\cup \{l_i: 2<i\le k \}$,
 and let $\C_1=\C\backslash (C_1\cup \{1\})$. Then
$\mid\C_1\mid= m^2-m -k> (m-1)^2$.
From  \cite{BT}(Theorem 10),
we have two columns sets $C_2$ and $C_3$
satisfying  $\mid C_2\mid=1$, $\mid C_3\mid=w+1-k$, and
$C_2$ and $C_3$ are not separable in the rows from $k+1$ to $w+2$.
Thus, we have $C_1\cup C_3$ is not separated from $C_2\cup \{1\}$.
 The proof is complete.\qed

\begin{theorem}
\label{4.4-2w} Suppose $A$ is a representation matrix of an
{\rm SHF}$(2+w;n,m,\{2,w\})$ with $w\ge2$ and $m \geq 2+w $, then $ n<
m^2-m$.
\end{theorem}
\noindent {\it Proof:} We use induction on $w$ to prove the theorem.

1. By Theorem~\ref{SHFM}, for $w=2$ this case satisfies.

2.  Assume the conclusion holds for $w=k-1, k\ge 3$.
 Suppose, for a contradiction, that  an SHF$(k+2;m^2-m+3,m,\{2,k\})$ exists with $A$ as  the representation matrix.
 Let $\C$ denote the set of columns of $A$. By removing the first row of $A$,
we obtain a $(k+1)\times (m^2-m+3)$ submatrix $B$. By
inductive hypothesis, there are two sets of columns $C_1$ and
$C_2$ in $B$ with $|C_1|= 2$ and $|C_2|= k -1$ which are not
separable. Now, we consider the same column sets $C_1$ and $C_2$
in $A$. Let $\C_1=\C\backslash C_1$.
If there exist two columns $l_1$ and $l_2$ satisfying $l_1\in \C_1$ and $l_2\in C_1$
such that $a_{1,l_1}=a_{1,l_2}$,
then $C_1$ is not separated from $C_2\cup\{l_1\}$,  a contradiction;
Now we have that for any two columns  $l_1\in \C_1$ and $l_2\in C_1$ satisfy $a_{1,l_1}\neq a_{1,l_2}$.
Thus, $\mid \{a_{1,i}: i\in \C_1\}\mid \le m-1$.
Similarly, we have $\mid \{a_{2,i}: i\in \C_1\}\mid \le m-1$.
By the pigeonhole principle, there is an
element $t$ in the first row such that $\lambda_t^{1}\ge \lceil \frac{n-2}{m-1}\rceil> m-1$.
So we have $d_{1,2}(t,z)\ge 2$ in $\C_1$. It contradicts Lemma \ref{4.3aabb}.
The proof is complete.\qed

We use Theorem \ref{4.4-2w} as the inductive hypothesis to prove the following theorem.
The proof method is similar to the Theorem \ref{WW}.
\begin{theorem}\label{4.5-w_1w_2}
  If there exists an {\rm SHF}$(w_1+w_2;n, m,\{w_1, w_2\})$ with $2\le w_1< w_2$ and $m \ge w_1+w_2$,
 then $ n< m^ 2-m$.
\end{theorem}

\noindent \textbf{Remark 4 :}
The best upper bound for  an SHF$(w_1+w_2; n,m,\{w_1,w_2\})$ with $m \geq w_1+w_2$  is $n \le m^2$ \cite{BT}.
When $w_1=1$ and $w_2\ge 2$, this  bound is tight.
When $w_2 > w_1 \ge 2$, we improve this bound from $n\le m^2$ to $n < m^2 -m$.

Now, we obtain the mainly conclusion of SHFs in the following.

\begin{theorem}\label{4.5-w_1w_t}
Suppose there exists an {\rm SHF}$(u;n, m,\{w_1, w_2,\ldots,w_t\})$ with $u = \sum_{i=1}^{t}w_i$. If $m\ge u$
  and $\{w_1, w_2,\ldots,w_t\}\notin \{ \{1,1,1\}, \{1,w\}\}$,
 then $ n< m^ 2-m$.
\end{theorem}

\noindent {\it Proof:} If  $t=2$, since $\{w_1, w_2,\} \neq \{1,w\}$ we have $ n< m^ 2-m$
by  Theorems \ref{WW} and \ref{4.5-w_1w_2}.
Now we assume $t\ge 3$.
 By the definition of an SHF,
we know that an SHF$(u;n, m,\{w_1, w_2,\ldots,w_t\})$ is also an SHF$(u;n, m,\{w, w^{\prime}\})$ with  $w=w_1+w_2$ and $w^{\prime}=\sum_{i=3}^{t}w_i$.
Since   $\{w_1, w_2,\ldots,w_t\}\neq  \{1,1,1\}$, we have $w\ge 2$ and $w^{\prime}\ge 2$.
By Theorems \ref{WW} and \ref{4.5-w_1w_2},  we have $ n< m^ 2-m$.\qed

\noindent \textbf{Remark 5:} The best upper bound for an SHF$(N;n,m,\{w_1,w_2,\ldots,w_t\})$
is $n\le rm^{\lceil\frac{N}{u-1}\rceil}$ $+(u-r)m^{\lfloor\frac{N}{u-1}\rfloor}$ where
$u=\sum_{i=1}^{t}w_i$, $1\le r\le u-1$ and  $N\equiv r \pmod{u-1}$ \cite{SGG}.
When $N=u$, we improve this bound from $n\le m^2+ (u-1)m$ to $n< m^2-m$.

\end {document}